\documentclass[10pt,twocolumn,letterpaper]{article}

\usepackage{cvpr}              

\usepackage{graphicx}
\usepackage{amsmath}
\usepackage{amssymb}
\usepackage{booktabs}

\DeclareMathOperator*{\argmin}{arg\,min}
\usepackage{enumitem}
\usepackage{multirow}
\usepackage[accsupp]{axessibility} 
\usepackage[ruled,vlined]{algorithm2e}

\usepackage[pagebackref,breaklinks,colorlinks]{hyperref}

\usepackage[capitalize]{cleveref}
\crefname{section}{Sec.}{Secs.}
\Crefname{section}{Section}{Sections}
\Crefname{table}{Table}{Tables}
\crefname{table}{Tab.}{Tabs.}

\def\cvprPaperID{10799} 
\def\confName{CVPR}
\def\confYear{2022}

\begin{document}

\title{
    HyperSegNAS: Bridging One-Shot Neural Architecture Search with \\3D Medical Image Segmentation using HyperNet 
}

\author{Cheng Peng\textsuperscript{1} \hspace{2mm} Andriy Myronenko\textsuperscript{2} \hspace{2mm} Ali Hatamizadeh\textsuperscript{2} \hspace{2mm} Vish Nath\textsuperscript{2} \hspace{2mm} Md Mahfuzur Rahman Siddiquee\textsuperscript{3} \\Yufan He\textsuperscript{2} \hspace{2mm} Daguang Xu\textsuperscript{2} \hspace{2mm} Rama Chellappa\textsuperscript{1} \hspace{2mm} Dong Yang\textsuperscript{2}\\
\textsuperscript{1}Johns Hopkins University \hspace{2mm}
\textsuperscript{2}NVIDIA \hspace{2mm} \textsuperscript{3}Arizona State University}
\maketitle

\begin{abstract}
   Semantic segmentation of 3D medical images is a challenging task due to the high variability of the shape and pattern of objects (such as organs or tumors).
   Given the recent success of deep learning in medical image segmentation, Neural Architecture Search (NAS) has been introduced to find high-performance 3D segmentation network architectures.
   However, because of the massive computational requirements of 3D data and the discrete optimization nature of architecture search, previous NAS methods require a long search time or necessary continuous relaxation, and commonly lead to sub-optimal network architectures.
   While one-shot NAS can potentially address these disadvantages, its application in the segmentation domain has not been well studied in the expansive multi-scale multi-path search space.
   To enable one-shot NAS for medical image segmentation, our method, named HyperSegNAS, introduces a HyperNet to assist super-net training by incorporating architecture topology information. Such a HyperNet can be removed once the super-net is trained and introduces no overhead during architecture search. We show that HyperSegNAS yields better performing and more intuitive architectures compared to the previous state-of-the-art (SOTA) segmentation networks; furthermore, it can quickly and accurately find good architecture candidates under different computing constraints. Our method is evaluated on public datasets from the Medical Segmentation Decathlon (MSD) challenge, and achieves SOTA performances.
\end{abstract}

\section{Introduction}
\label{sec:intro}
\begin{figure}[!htb]
    \setlength{\abovecaptionskip}{3pt}
    \setlength{\tabcolsep}{2pt}
    \vspace{-1em}
    \begin{tabular}[b]{c}
        \begin{subfigure}[b]{\linewidth}
            \includegraphics[width=\textwidth]{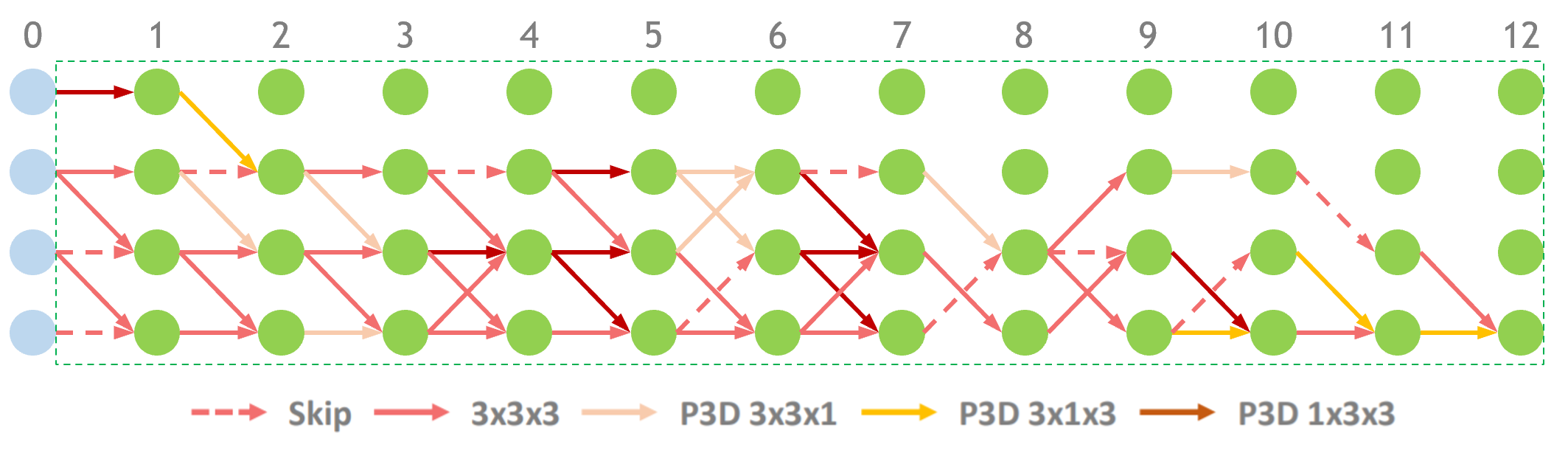}
            \caption{DiNTS \cite{DBLP:conf/cvpr/He0RZX21} Searched Architecture, 5,067~MB}
            \label{dints}
        \end{subfigure} \\
        \begin{subfigure}[b]{\linewidth}
            \includegraphics[width=\textwidth]{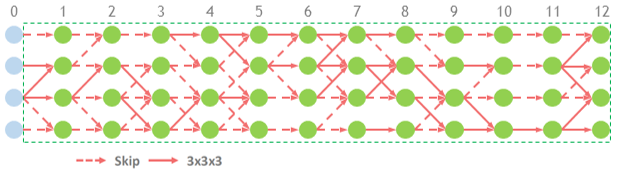}
            \caption{HyperSegNAS Searched Architecture, 5,358~MB}
            \label{ours}
        \end{subfigure} \\
        \begin{subfigure}[b]{\linewidth}
            \includegraphics[width=\textwidth]{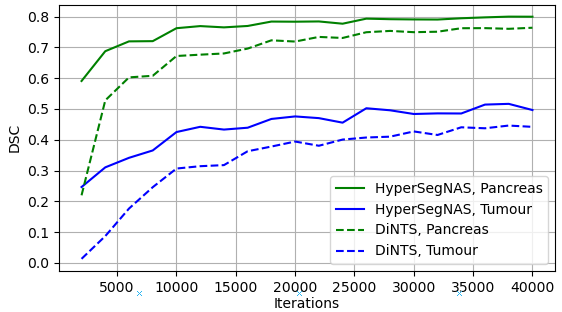}
            \caption{Five-fold cross-validation for \ref{dints} and \ref{ours}}
            \label{perf}
        \end{subfigure} \\
    \end{tabular}
    \caption{Architectures found from DiNTS~\cite{DBLP:conf/cvpr/He0RZX21} (\ref{dints}) and our proposed HyperSegNAS (\ref{ours}) based on the Pancreas dataset in MSD \cite{DBLP:journals/corr/abs-2106-05735}. Under similar compute costs, HyperSegNAS' architecture uses more skip connections and ensure features of multiple scales are propagated for predictions. Our architecture performs significantly better than DiNTS, as shown in \ref{perf}.}
    \label{fig:intro}
    \vspace{-2em}
\end{figure}
Automated medical image segmentation is an active research area with many clinical applications including anatomical analysis~\cite{monteiro2020multiclass} and disease diagnosis~\cite{Harmon2020}. Medical image segmentation remains challenging despite advances in deep learning methods. To address the large variability of target objects, the limited number of data samples, and the higher dimensional and high resolution nature of 3D images, researchers have spent a great effort in designing efficient neural architectures under various computing settings~\cite{DBLP:conf/miccai/CicekALBR16,DBLP:conf/miccai/RonnebergerFB15,DBLP:conf/3dim/MilletariNA16,DBLP:conf/miccai/LiuXZPGMWJCC18,DBLP:journals/tmi/DolzGYLDA19}.

Neural Architecture Search (NAS) has surfaced in response to handcrafted approaches and promises to find well performing architecture through automated algorithms. While initially focusing on classification~\cite{DBLP:conf/eccv/YuJLBKTHSPL20,DBLP:conf/iclr/CaiGWZH20,DBLP:conf/iclr/XuX0CQ0X20,DBLP:conf/iclr/LiuSY19,DBLP:conf/iclr/ZophL17}, NAS has been also introduced to the segmentation domain in recent years~\cite{DBLP:conf/miccai/YanJSZ20,DBLP:conf/wacv/XiaLYCYZXYR20,DBLP:journals/access/WengZLQ19,DBLP:conf/miccai/KimKLBKCYK19,DBLP:conf/cvpr/He0RZX21,DBLP:conf/cvpr/LiuCSAHY019,DBLP:conf/cvpr/YuYRBZYX20,DBLP:conf/3dim/ZhuLYYX19}. Segmentation networks usually preserve features from multiple scales and aggregate them together to accurately segment objects with different sizes. Accordingly, the search space for multi-scale, multi-path architectures can be very complex, as the feasible network configurations increase exponentially with scale. C2FNAS~\cite{DBLP:conf/cvpr/YuYRBZYX20} needs close to \emph{one GPU year} to search for a single 3D segmentation architecture based on the Evolutionary Algorithm (EA), despite building on a U-shape network architecture. On the other hand, DiNTS~\cite{DBLP:conf/cvpr/He0RZX21} relaxes the architecture search problem from a discrete formulation to a continuous and differentiable one, thus it greatly speeds up the searching process. Such a relaxation, however, may lead to (1) an optimization gap when the continuous architecture/edge weights are discretized for deployment~\cite{DBLP:conf/iccv/ChenXW019,DBLP:journals/pr/Tian0XJY21} and (2) possibly infeasible architectures, which require ad-hoc logic to be handled. 

Beyond finding the best performing architecture, balancing other computing constraints to the search process is another important aspect of NAS and has led to popularity in methods like one-shot NAS~\cite{DBLP:conf/nips/SaxenaV16,DBLP:conf/icml/PhamGZLD18,DBLP:conf/icml/BenderKZVL18,DBLP:conf/iclr/BrockLRW18,DBLP:conf/iclr/CaiZH19,DBLP:conf/cvpr/HuangC21,DBLP:conf/eccv/GuoZMHLWS20,DBLP:conf/cvpr/YouHYWQZ20,DBLP:conf/cvpr/ChenFL21,DBLP:journals/corr/abs-1911-13053,DBLP:conf/eccv/YuJLBKTHSPL20,DBLP:conf/iclr/CaiGWZH20}, where all sub-nets are trained with shared parameters under a large super-net. This aspect is especially important for medical imaging segmentation task for which memory usage is a major challenge when searching for the optimal architecture across various devices and resource constraints. Once training is completed, the shared parameters can be used to evaluate the possible sub-nets and select the best performers given the specified constraints. So far, one-shot approaches have not been applied to the 3D segmentation's search space yet. As there is no clear training strategies, such as the unilateral augmented (UA) principle~\cite{DBLP:conf/cvpr/SuY00Z021} used in progressive shrinking~\cite{DBLP:conf/iclr/CaiGWZH20} and the Sandwich Rule~\cite{DBLP:conf/eccv/YuJLBKTHSPL20,DBLP:conf/iccv/YuH19}, in a multi-scale, multi-path search space\cite{DBLP:conf/cvpr/LiuCSAHY019}, we find that the one-shot training scheme with randomly sampled architecture topology frequently leads to sub-optimal performance.



We propose HyperSegNAS, which follows the one-shot NAS approach and seeks to address its issues in the segmentation space. Within HyperSegNAS, we propose a novel Meta-Assistant Network (MAN) to improve super-net training. Specifically, both the sampled architecture and the input image are fed to MAN. MAN then dynamically modifies the shared weights during the training process based on deploying architectures. We show that such an \emph{architecture topology-aware} training method significantly improves sub-net performances. When training is completed, an annealing process is used to gradually remove MAN from the main network. Even after removing MAN, the sub-net performances remain high and can be efficiently evaluated.

Benefiting from the one-shot paradigm, HyperSegNAS has great advantages in multi-fold. Firstly, HyperSegNAS is much more efficient at evaluating over \emph{many architectures} under different computing constraints, as the shared parameters only need to be trained once. In comparison, approaches in ~\cite{DBLP:conf/cvpr/He0RZX21,DBLP:conf/cvpr/YuYRBZYX20} need to search from scratch to obtain each single architecture. Secondly, HyperSegNAS evaluates on discrete architectures, which eliminates the discretization gap and infeasible configurations that arise in differentiable NAS. Finally, HyperSegNAS can accurately fit to the given constraints due to the decoupling between training and searching. As demonstrated in Figures~\ref{dints} and~\ref{ours}, our search algorithm results in a very different architecture as compared to DiNTS's under similar computing constraints. Not only is our architecture significantly better performing, as shown in Fig. \ref{perf}, it also follows conventional intuitions in propagating multi-scale features through the use of skip connections. Finally, HyperSegNAS achieves new SOTA performance on multiple tasks in the Medical Segmentation Decathlon (MSD) challenge~\cite{DBLP:journals/corr/abs-2106-05735} using similar computing budget as DiNTS.

In summary, our main contributions are listed as follows:
\begin{itemize}
    \item We propose HyperSegNAS, a one-shot NAS approach at finding optimal 3D segmentation architectures. HyperSegNAS can search for efficient architectures that accurately adapt to different computing constraints.
    \item We propose a HyperNet structure called Meta-Assistant Network (MAN) to address the large search space of segmentation networks by incorporating relevant meta information to NAS; MAN can be removed after training, and does not add computing overhead for searching and re-training.
    \item We achieve better results on both low- and high-compute architectures compared to DiNTS and new SOTA performances in multiple tasks in MSD.
\end{itemize}





\section{Related Work}
\subsection{Medical Image Segmentation}

Medical image segmentation has been studied and developed for decades in many medical imaging applications.
Particularly, deep learning based-algorithms have been widely adopted in segmentation applications in recent years.
Among the deep neural network models, U-Net~\cite{DBLP:conf/miccai/RonnebergerFB15,DBLP:conf/miccai/CicekALBR16} is a seminal work in this area due to its effectiveness and structural simplicity for processing multi-scale image features.
Various improvements have been introduced subsequently~\cite{DBLP:conf/3dim/MilletariNA16,DBLP:journals/tmi/ZhouSTL20,DBLP:journals/tmi/LiCQDFH18,Isensee2021}, which include adding components like residual blocks or dense blocks and combining 2D and 3D U-Nets.
On the other hand, nn-UNet~\cite{Isensee2021} ensembles predictions from multiple existing 2D/3D U-Net models, and focuses on finding the correct configurations/hyper-parameters to improve the performance.
There are also other existing architectures without U-shape structure, e.g. AH-Net~\cite{DBLP:conf/miccai/LiuXZPGMWJCC18}, HDenseNet~\cite{li2018hdenseunet}, HyperDenseNet~\cite{DBLP:journals/tmi/DolzGYLDA19}, etc.
A common theme across these works is to ensure features of various dimension and resolution are preserved while maintaining manageable computational cost; therefore, the use of deep learning modules (e.g., convolution, skip connections) has been a major area of exploration for manually designed architectures.

\subsection{Neural Architecture Search (NAS)}

Manual architecture designs have resulted in steady performance improvements across vision tasks; however, they iterate gradually and with great recurring cost. NAS proposes to automate architecture design in a performance-driven way, and has received consideration attention. NAS has been frequently applied on image recognition tasks~\cite{DBLP:conf/eccv/YuJLBKTHSPL20,DBLP:conf/iclr/CaiGWZH20,DBLP:conf/iclr/XuX0CQ0X20,DBLP:conf/iclr/LiuSY19,DBLP:conf/iclr/ZophL17}. On a high level, evolutionary algorithm (EA)~\cite{DBLP:conf/aaai/RealAHL19} and reinforcement learning (RL)~\cite{DBLP:conf/iclr/ZophL17} are used to perform discrete search/optimization, differentiable relaxation~\cite{DBLP:conf/iclr/LiuSY19} on connections and operations allow search algorithms to leverage gradient descent optimization. Finally, training an architecture performance estimator, e.g. one-shot NAS~\cite{DBLP:conf/nips/SaxenaV16,DBLP:conf/icml/PhamGZLD18,DBLP:conf/icml/BenderKZVL18,DBLP:conf/iclr/BrockLRW18,DBLP:conf/iclr/CaiZH19,DBLP:conf/cvpr/HuangC21,DBLP:conf/eccv/GuoZMHLWS20,DBLP:conf/cvpr/YouHYWQZ20,DBLP:conf/cvpr/ChenFL21,DBLP:journals/corr/abs-1911-13053,DBLP:conf/eccv/YuJLBKTHSPL20,DBLP:conf/iclr/CaiGWZH20}, which our method is based on, is another popular approach as it disentangles the training and searching processes. While previous methods~\cite{DBLP:conf/iclr/BrockLRW18,DBLP:conf/iclr/ZhangRU19} have used HyperNets to directly estimate sub-net performances, HyperSegNAS focuses on accelerating feature learning through an assistant HyperNet, which does not introduce an overhead in the search process.

More recently, NAS was explored in medical image segmentation~\cite{DBLP:conf/miccai/YanJSZ20,DBLP:conf/wacv/XiaLYCYZXYR20,DBLP:conf/cvpr/He0RZX21}. As 3D segmentation architectures need to process and preserve multi-resolution features, they are much more computationally intense than 2D classification architectures. Early attempts~\cite{DBLP:journals/access/WengZLQ19,DBLP:conf/3dim/ZhuLYYX19,DBLP:conf/miccai/KimKLBKCYK19,DBLP:conf/cvpr/YuYRBZYX20} build on the prior architecture of U-Net, and modify edge operations and connections while preserving the overall U-shape structure. To go beyond a U-Net based structure while remaining computationally feasible, DiNTS~\cite{DBLP:conf/cvpr/He0RZX21} applies a differentiable NAS formulation similar to ~\cite{DBLP:conf/iclr/LiuSY19,DBLP:conf/iclr/XuX0CQ0X20} on a segmentation search space first proposed by Auto-DeepLab~\cite{DBLP:conf/cvpr/LiuCSAHY019}. Due to the efficiency introduces in differentiable NAS, DiNTS~\cite{DBLP:conf/cvpr/He0RZX21} can feasibly search for architectures that fit under different computing budgets, and is the current best performer in the MSD~\cite{DBLP:journals/corr/abs-2106-05735} challenge.

\begin{figure*}[!htb]
    \setlength{\abovecaptionskip}{3pt}
    \setlength{\tabcolsep}{0.5pt}
    \begin{tabular}[b]{cc}
        \begin{subfigure}[b]{0.748\linewidth}
           \includegraphics[width=\textwidth]{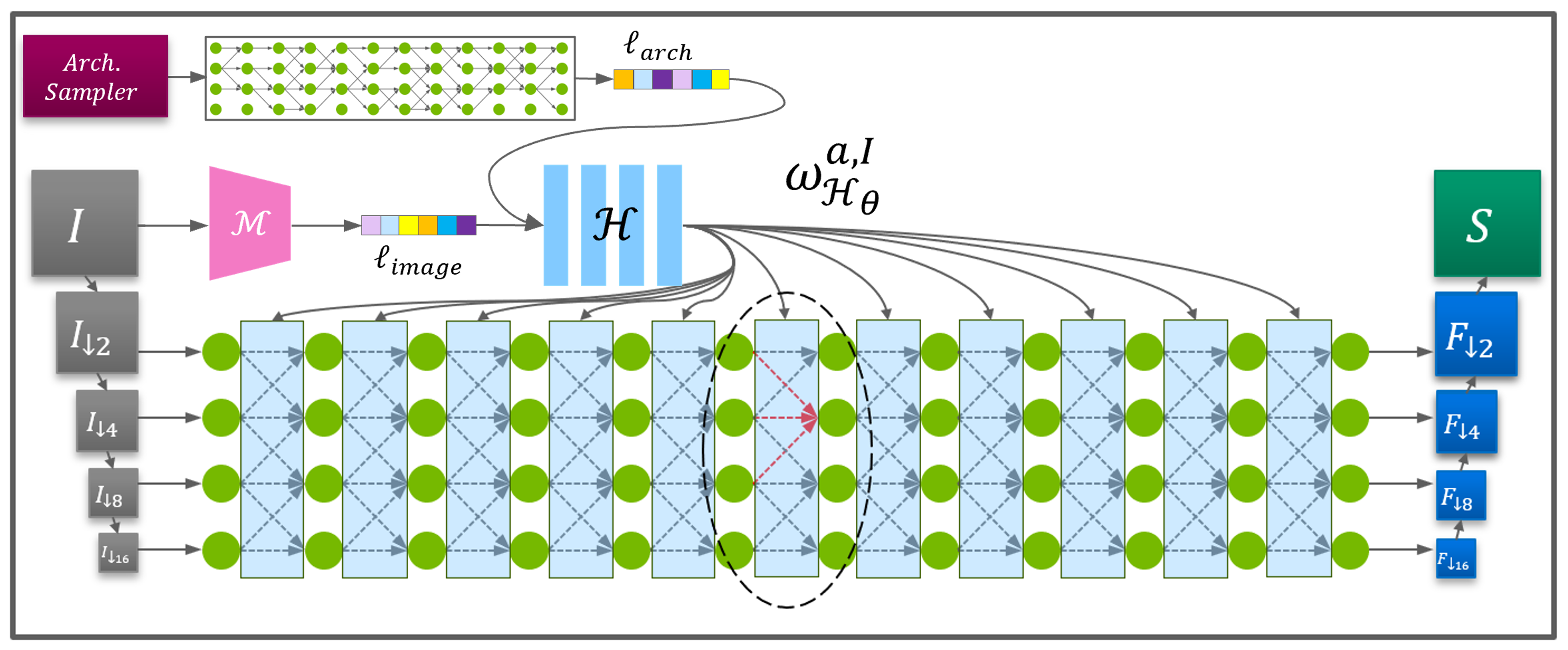}
            \caption{The training pipeline of HyperSegNet. Details on the selected edges (red dash arrows) are expanded in \ref{detail_pipeline}.}
            \label{main_pipeline}
        \end{subfigure} &
        \begin{subfigure}[b]{0.255\linewidth}
            \includegraphics[width=\textwidth]{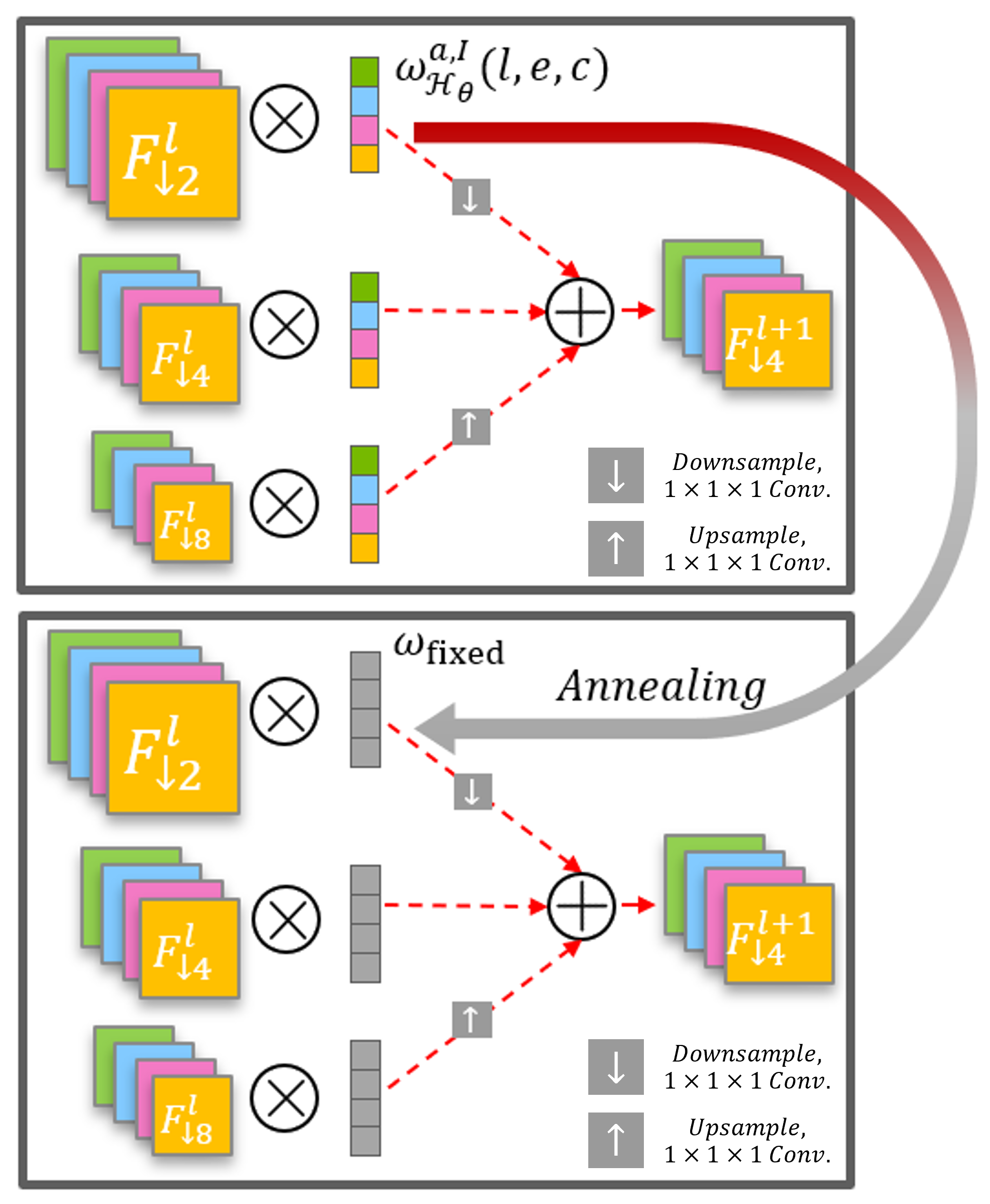}
            \caption{Detailed view of \ref{main_pipeline}}
            \label{detail_pipeline}
        \end{subfigure}
    \end{tabular}
    \caption{HyperSegNet introduces an additional HyperNet $\mathcal{H}$ on top of the segmentation search space. When training the super-net, $\mathcal{H}$ introduces additional channel-wise weights, which depend on the input image and the sampled architecture. The circled block in~\ref{main_pipeline} is expanded in~\ref{detail_pipeline} for a detailed view of the feature re-weighting process. An annealing process is introduced to gradually replace $\omega^{a,I}_{\mathcal{H}_{\theta}}$ with $\omega_{\textrm{fixed}}$ and removes $\mathcal{H}$ from the inference process. As such, efficiency and better estimated performance are achieved.} 
    \label{fig:pipeline}
    \vspace{-1em}
\end{figure*}

\section{Method}

\subsection{Network Search Space}
\textbf{Topology.} We follow previous works~\cite{DBLP:conf/cvpr/LiuCSAHY019,DBLP:conf/cvpr/He0RZX21} in defining the search space for 3D segmentation architectures. As shown in Fig.~\ref{fig:pipeline}, we down-sample the input image $I$ to generate multi-scale $I_{\downarrow r}$, where $r$ denotes the down-sampling factor. Overall, the search space has $L=12$ layers and contains features of $|r|=4$ scales, denoted as $F^{l}_{\downarrow r}$, at every layer, where $l\in \{1,2,...,L\}$. There are $E=10$ possible edges between layers; each edge contains a basic operation and an up-sampling/down-sampling/identity operation depends on the 
output feature scale. When multiple edges point to the same feature scale, their outputs are summed and normalized. As shown in~\cite{DBLP:conf/cvpr/LiuCSAHY019,DBLP:conf/cvpr/He0RZX21}, this search space is very general in topology and contains many major segmentation architectures (like U-shape networks). 

\textbf{Edge Operations.} For each edge in the search space, a basic operation is chosen from a pre-defined operation set. As the complexity in topology is already high, we take a minimalist approach at choosing the basic operation set, which includes skip connection and $3 \times 3 \times 3$ 3D convolution. Compared to~\cite{DBLP:conf/cvpr/He0RZX21}, we omit three pseudo 3D operators (P3D). Such a design choice (1) reduces the search space complexity and allows for better clarity on network topology, (2) still contains the majority of popular medical segmentation architectures, e.g. 3D U-Net and its variants. 

\subsection{HyperSegNAS}
\textbf{Background.}
 The goal of NAS is to find the best performing sub-network $a$ from the search space $\mathcal{A}$ under a given computing constraint $\mathcal{C}$, e.g., peak GPU memory at training, as defined by its minimum validation loss 
 \begin{align}\label{eq:val}
 \argmin_{a}\mathcal{L}_{\textrm{val}}(a, \omega^*_a), \forall a \in \mathcal{A} \mid Cost(a)<\mathcal{C},
 \end{align}
 where $\omega^*_a$ is the optimal weights of $a$ after training. Since $\omega^*_a$ is generally very costly to obtain from a full training process, the \textbf{one-shot NAS} defines a super-net with weights $\omega$, which covers the entire search space. After $\omega$ is trained, it is used as a performance estimator of $a$ based on the sub-set weights $\omega_a\in\omega$. This approach means that all sub-nets are evaluated based on shared weights; the training objective is formulated as
 \begin{align}\label{eq:train}
 \argmin_{\omega}\frac{\sum\mathcal{L}_{\textrm{train}}(a, \omega_a)}{|\mathcal{A}|}, \forall a \in \mathcal{A}.
 \end{align}
Clearly, it is infeasible to enumerate all architectures for inference and average their gradients; in practice, one or more architectures are sampled at each training step for local averaging. After training, the best architecture $a^*$ is then selected by evaluating
 \begin{align}\label{eq:shared_val}
 \argmin_{a}\mathcal{L}_{\textrm{val}}(a, \omega_a), \forall a \in \mathcal{A} \mid Cost(a)<\mathcal{C}.
 \end{align}
 The one-shot formulation reduces the complexity of obtaining $\omega^*_{a}$ in Eq.~\ref{eq:val} through weight sharing and can directly evaluate discrete architectures; however, it also has limitations. When the search space $\mathcal{A}$ is large, optimizing Eq.~\ref{eq:train} becomes very different from optimizing individual sub-nets. The \emph{multi-scale, multi-path} architectures in our segmentation search space also introduces challenges in training the super-net. Unlike search dimensions like kernel size and channel size~\cite{DBLP:conf/iclr/CaiGWZH20,DBLP:conf/eccv/YuJLBKTHSPL20,DBLP:conf/iccv/YuH19}, a multi-path architecture topology does not have a self-evident order and cannot be trained in a progressive manner. We find that training the super-net from randomly sampled topology leads to sub-optimal performance evaluations on sub-nets. As such, we hypothesize that due to the lack of good training strategies and a large search space, these super-net features tend to have high redundancy and suffer in accuracy.
 



\textbf{Meta-Assistant Network.}
To address the problem discussed above, we propose a Meta-Assistant Network (MAN), which is architecture topology-aware. MAN relaxes the formulation in Eq.~\ref{eq:train} by providing extra degrees of freedom on $\omega$ to improve training, i.e. changing $\omega$ based on meta information about the problem. In particular, we focus on two pieces of meta information: the architecture $a$ and the input image $I$. As shown in Fig.~\ref{fig:pipeline}, MAN consists of an architecture sampler, an image encoder $\mathcal{M}$, and a HyperNet $\mathcal{H}_{\theta}$, where $\theta$ denotes $\mathcal{H}$'s parameters. We represent the sampled architecture $a\in \mathbb{N}^{L\times E\times O}$ as a matrix, where $O=2$ denotes the cardinality of the edge operations. All activated edges and cell operations in the respective layers are labeled as 1, and 0 otherwise. 

The conventional implementations of HyperNet~\cite{DBLP:conf/iclr/HaDL17,DBLP:conf/iclr/BrockLRW18} directly output network weights $\omega$ based on relevant information. However, the total number of parameters $|\omega|$ is exceedingly large and costly to generate. Instead, MAN generates channel-wise scalar weights, as demonstrated in Fig.~\ref{detail_pipeline}. 
More precisely, we denote every convolution kernel in the super-net as $\omega(l,e,c) \in \mathbb{R}^{k\times k}$, where $k$ is the kernel size. Respectively, $l\in \{1,2,...,L\}$, $e\in \{1,2,...,E\}$, and $c$ are the layer, edge, and channel index. 
For every training iteration, an architecture $a$ and a training image $I$ is sampled. We obtain an image vector $l_{\textrm{image}}$ from $\mathcal{M}(I)$, and an architecture vector $l_{\textrm{arch}}$ by flattening the architecture matrix representation. These vectors are then concatenated as the input to $\mathcal{H}$ to obtain $\omega^{a,I}_{\mathcal{H}_\theta} = \mathcal{H}_\theta(l_{\textrm{arch}}\oplus l_{\textrm{image}})$.
Finally, the channel-wise weights $\omega^{a,I}_{\mathcal{H}_\theta}$ are multiplied with the kernels in the super-net and transform training of Eq.~\ref{eq:train} into

 \begin{align}\label{eq:hypernet}
 \argmin_{\omega,\theta}\frac{\sum\mathcal{L}_{\textrm{train}}(a, \omega^{a,I}_{\mathcal{H}_\theta}(l,e,c)*\omega_a(l,e,c))}{|\mathcal{A}|}.
 \end{align}

Under this formulation, individual sub-networks gain additional capacity to optimize their own configurations based on a given image. We observe that this setting significantly improves performance.

\textbf{HyperNet Annealing.} While MAN improves sub-network learning, it also leads to potentially biased evaluations due to the introduction of a new network component and the overall increased representational capacity. This can affect the correlation between sub-network's estimated performance from shared weights and true performance in deployment, especially for smaller architectures. To obtain shared weights that are independent of MAN and ensure fairness in searching, HyperSegNAS consists of a second stage to remove MAN for evaluation. Specifically, we introduce a constant channel-wise scalar weight $\omega_{\textrm{fixed}}=0.5$ and an annealing process to gradually replace $\omega^{a,I}_{\mathcal{H}_\theta}$ with $\omega_{\textrm{fixed}}$. Such a process can be expressed as:
\vspace{-0.5em}
 \begin{align}\label{eq:anneal}
 \argmin_{\omega,\theta}\frac{\sum\mathcal{L}_{\textrm{train}}(a, (\lambda\omega_{\textrm{fixed}}+(1-\lambda)\omega^{a,I}_{\mathcal{H}_\theta})*\omega_a))}{|\mathcal{A}|},
 \end{align}
where $\lambda$ is the annealing temperature that gradually moves from 0 to 1. We find that a linear annealing schedule works well and allows $\omega$ to maintain reasonable performance after MAN is removed, thus achieving an overall improvement in estimating performances based on $\omega$ alone.

\textbf{Architecture Search.} After $\omega$ is properly trained, we evaluate different architectures and select the best performing candidates that fit under the desired computing constraints. While HyperSegNAS yields observably better estimated sub-network performances, these estimations are still slightly worse in comparison to that of $\omega^*_a$ from training the architecture from scratch, as is generally observed in one-shot literature~\cite{DBLP:conf/iclr/CaiGWZH20,Bu_2021_CVPR}. More competitive performances can be reached by applying a few architecture-specific finetuning steps on $\omega$.

As doing so across all architectures is very expensive, we deploy a coarse-to-fine search algorithm, which can be described in three parts. 
\begin{enumerate}
\item We sample a sufficient number of candidate architectures, i.e. architectures that fit under a given computing constraint $\sigma$, to form a valid search space $\mathcal{A}_{\sigma}$.

\item We evaluate all candidates in $\mathcal{A}_{\sigma}$ based on the super-net and Eq.~\ref{eq:shared_val}, and select the top $N$ performers.

\item We fine-tune the $N$ architectures for a few iterations, evaluate their fine-tuned performances, and select the best performer as the final architecture.

\end{enumerate}

While any form of computing constraint, e.g., latency, is applicable to HyperSegNAS, we follow~\cite{DBLP:conf/cvpr/He0RZX21} in using training time GPU memory cost as the  computing constraint. Due to the decoupling of training and searching in Eq.~\ref{eq:train} and Eq.~\ref{eq:shared_val}, HyperSegNAS can precisely calculate the computing budgets of the evaluated model. Specifically, during the sampling stage in our search algorithm, HyperSegNAS can directly measure training memory cost by performing training for few iterations. This is a distinct advantage of HyperSegNAS compared to differentiable NAS methods like DiNTS~\cite{DBLP:conf/cvpr/He0RZX21}, as training memory cost is hard to be analytically formulated and predicted due to various low-level optimization and overheads at runtime~\cite{DBLP:conf/sigsoft/GaoLZLZLY20}. In fact, incorporating a differentiable analytical formulation of memory usage in DiNTS~\cite{DBLP:conf/cvpr/He0RZX21} leads to not only imprecise cost estimations, but also potentially biased architectures, e.g., as shown in Fig.~\ref{fig:intro}, due to the discretization gap.

\section{Experiments}
\begin{figure}[!htb]
    \centering
      \includegraphics[width=\linewidth]{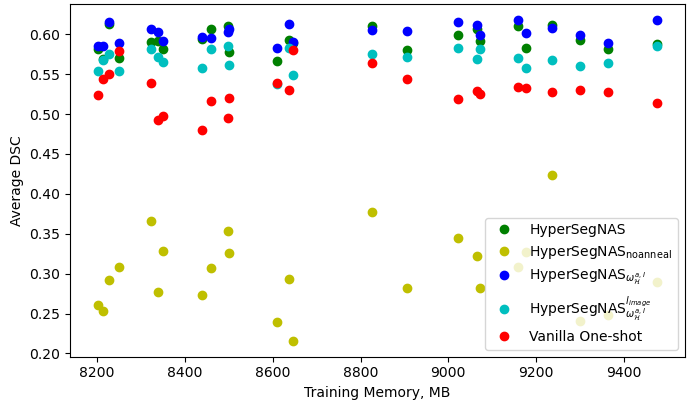}
    \caption{Visualization of the average Dice-Sørensen (DSC) scores on twenty-five evaluated architectures in Table~\ref{tab:ablation} sorted based on its training memory cost.}
    \label{fig:ablation}
    \vspace{-1em}
\end{figure}
\begin{table}[!htb]
    \setlength{\tabcolsep}{1.5pt}
    \centering
    \begin{tabular}[b]{c}
        \begin{tabular}{l|ccc|c}
        \toprule
        \multicolumn{1}{c|}{Method} & Pancreas & Tumor & Avg. & Inf. Speed \\
        \midrule
        Vanilla One-Shot 
        & 71.93
        & 33.96
        & 52.92
        & 44.60s\\
        \midrule
        $\textrm{HyperSegNAS}^{l_{I}}_{\omega^{a,I}_{\mathcal{H}}}$
        & 75.64
        & 37.42
        & 56.53
        & 67.91s\\
        \midrule
        $\textrm{HyperSegNAS}_{\omega^{a,I}_{\mathcal{H}}}$
        & 75.90
        & \textbf{44.36}
        & \textbf{60.13}
        & 69.41s\\
        \midrule
        $\textrm{HyperSegNAS}_{\textrm{noanneal}}$
        & 32.15
        & 28.17
        & 30.16
        & 46.89s\\
        \midrule
        HyperSegNAS 
        & \textbf{76.31}
        & 41.99
        & 59.15
        & 47.27s\\
        \bottomrule
        \end{tabular}\\
    \end{tabular}
    \caption{Ablation study on different implementations of HyperSegNAS. Twenty-five architectures are randomly sampled and evaluated on the Pancreas dataset. Dice-Sørensen (DSC) Scores on respective labels and segmentation inference speed per 3D volume (in seconds) are reported.}
    \label{tab:ablation}
    \vspace{-1.5em}
\end{table}
\textbf{Dataset.} We follow the settings in~\cite{DBLP:conf/cvpr/He0RZX21,DBLP:conf/cvpr/YuYRBZYX20}. Specifically, our search algorithm is performed on the MSD Pancreas dataset, which has been shown by previous work to lead to generally performant architectures. We resample 281 provided scans to a voxel resolution of $1.0 \times 1.0 \times 1.0 mm^3$, use 225 scans for training, and 56 for validation. 

\textbf{Implementation Details.} 
We perform architecture search on the same topological search space as in~\cite{DBLP:conf/cvpr/He0RZX21}. For details, the channel size $c$ for features of scales $r=\{1,2,3,4\}$ are $\{32,64,128,256\}$; overall, there are a total number of $c_{tot}=27,648$ channels in the search space. During training, patches of size $96\times96\times96$ are sampled and augmented with rotation and flip. The image encoder $\mathcal{M}$ consists of six 3D convolutional layer with Instance Normalization~\cite{DBLP:journals/corr/UlyanovVL16} and ReLU activation, and outputs $l_{\textrm{image}}\in \mathbb{R}^{256}$. The HyperNet $\mathcal{H}$ consists of five fully connected layers with ReLU activations in between and a Sigmoid activation at the end; it outputs $\omega^{a,I}_{\mathcal{H}_\theta} \in \mathbb{R}^{c_{tot}}$. For every sampled architecture $a$, the inactivated portion of $\omega^{a,I}_{\mathcal{H}_\theta}$ is masked off from back-propagation. We use a combination of dice and cross-entropy loss as the training objective function $\mathcal{L}_{train}$. To speed up the evaluation during the search stage, we sample the foreground regions with a maximum size of $192\times192\times192\ mm^3$ and evaluate them patch-wise. The architecture sampling algorithm is implemented with several heuristics to eliminate undesirable configurations. We ensure that at least one feasible edge is sampled per layer so that the architecture always has $L$ layers. This prevents early termination and an excessive number of very small architectures. Feasible edges and operations are sampled with uniform likelihood. Super-net training is performed on eight NVIDIA V100 GPUs with a batch size of of eight and 160,000 iterations; HyperNet is then annealed over 20,000 iterations. 

Since HyperSegNAS can fit to cost constraint with accuracy, we constraint our searched architectures to have similar training memory costs as the three architectures found in~\cite{DBLP:conf/cvpr/He0RZX21}. Specifically, we sample two thousand architectures that cost within 300MB of DiNTS's respective architecture to form a $\mathcal{A}_{\sigma}$, and select the top $N=10$ architectures for quick fine-tuning with 5,000 iterations. Training memory is directly measured by monitoring the NVIDIA System Management Interface ($\emph{nvidia-smi}$ command). Overall, HyperSegNAS takes 32 GPU days to train the super-net, and 4 GPU days to search under one computing constraint. This is much faster than C2FNAS~\cite{DBLP:conf/cvpr/YuYRBZYX20}, which takes 333 GPU days to search for one architecture. Compared to DiNTS~\cite{DBLP:conf/cvpr/He0RZX21}, which takes 5.8 GPU days to search for one architecture, our method has a front-loaded cost in super-net training, but is faster at adapting to different computing constraints once super-net weights are obtained. Please refer to the Supplemental Material for more algorithm/implementation details.

\begin{figure}[!htb]
    \centering
      \includegraphics[width=\linewidth]{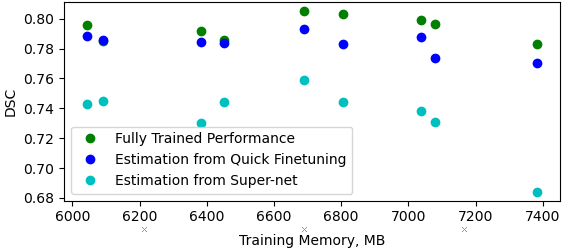}
    \caption{Improved ranking of architectures based on fine-tuned weights. The Kendall-Tau correlation between estimations from fine-tuned super-net weights and fully trained performances is $\tau=0.55$ , and $\tau=0.22$ based on using non fine-tuned super-net weights. Spearman's rank correlations are $r_s=0.7$ and $r_s=0.4$, respectively.}
    \label{fig:correlation}
    \vspace{-1em}
\end{figure}

\subsection{Ablation Study}
We evaluate the effectiveness of HyperSegNAS against alternative implementations. Specifically, we compare the proposed pipeline with the following implementations using the same training hyper-parameters:
\begin{enumerate}[label=\Alph*),itemsep=1 pt]
    \item Vanilla one-shot: The training scheme used in most one-shot methods, e.g.~\cite{DBLP:conf/eccv/YuJLBKTHSPL20,DBLP:conf/iclr/CaiGWZH20}, where the architecture is sampled and trained with shared weights.
    \item $\textrm{HyperSegNAS}^{l_{I}}_{\omega^{I}_{\mathcal{H}_{\theta}}}$: A HyperSegNAS implementation where the image vector $l_\textrm{image}$ is the only input to $\mathcal{H}$ and $\mathcal{H}$ is not removed.
    \item $\textrm{HyperSegNAS}_{\omega^{a,I}_{\mathcal{H}}}$: A HyperSegNAS implementation using both $l_{\textrm{image}}$ and $l_{\textrm{arch}}$ as input to $\mathcal{H}$; $\mathcal{H}$ is not removed.
    \item $\textrm{HyperSegNAS}_{\textrm{noanneal}}$: A HyperSegNAS implementation where $\mathcal{H}$ is removed without annealing.
\end{enumerate}
As shown in Figure~\ref{fig:ablation} and Table~\ref{tab:ablation}, the traditional one-shot training scheme leads to sub-optimal performance in our segmentation search space. The shared weights need to implicitly accommodate all randomly selected architectures; as the features are generally applicable to random architectures, their capacity is limited.  

In comparison, the addition of $\mathcal{H}$ helps generally improve performances across sampled architectures. Specifically, while $\textrm{HyperSegNAS}^{l_{I}}_{\omega^{a,I}_{\mathcal{H}}}$ shows some performance improvements, we observe significant improvements from applying both the image and architecture vectors in $\textrm{HyperSegNAS}_{\omega^{a,I}_{\mathcal{H}}}$, demonstrating the utility of each piece of meta-information in accelerating learning. While annealing $\mathcal{H}$ away leads to some performance drop off, as $\omega$ is now not architecture or input specific, HyperSegNAS still outperforms Vanilla one-shot by a wide margin. If annealing is not applied and $\mathcal{H}$ is simply removed from inference, performances are severely degraded. 


\subsection{Method Analysis}

\textbf{Search Correlation.} 
As HyperSegNAS explicitly evaluates the performances of different architectures based on shared fine-tuned weights, we can measure the Kendall-Tau correlation~\cite{kendall1938new} of such performance estimations with the performances of these architectures trained from scratch. As shown Fig.~\ref{fig:correlation}, we observe a moderate amount of correlation when shared weights are used for estimations, and a stronger correlation when fine-tuned weights are used. 

\begin{figure}[!htb]
    \setlength{\abovecaptionskip}{3pt}
    \setlength{\tabcolsep}{2pt}
    \vspace{-0.5em}
    \begin{tabular}[b]{c}
        \begin{subfigure}[b]{\linewidth}
            \includegraphics[width=\textwidth]{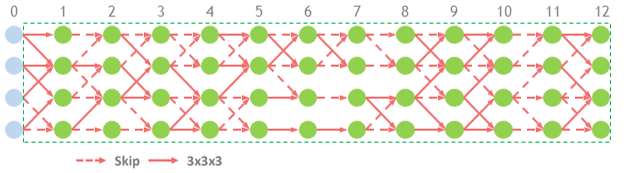}
            \caption{Searched Architecture, 7,168~MB}
        \end{subfigure} \\
        \begin{subfigure}[b]{\linewidth}
            \includegraphics[width=\textwidth]{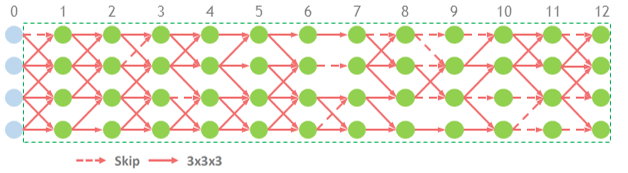}
            \caption{Searched Architecture, 9,173~MB}
        \end{subfigure} \\
    \end{tabular}
    \caption{Searched architectures from HyperSegNAS with various compute constraints.}
    \label{fig:structures}
    \vspace{-1em}
\end{figure}

\textbf{Searched Architectures.} Besides the low-cost architecture shown in Fig.~\ref{fig:intro}, we visualize the other two searched architectures that use comparable training memory to~\cite{DBLP:conf/cvpr/He0RZX21} in Fig.~\ref{fig:structures}. The training memory cost and performances based on five-fold cross-validation are presented in Table~\ref{tab:model_compare}. 

We make several observations. Firstly, under limited computational resources, e.g. at around 5GB of GPU training memory, the network architecture has a significant role on the final performance. This is reasonable as high compute architectures in the same search space tend towards homogeneity, i.e. the super-net, while low compute architectures have more variety. As such, we can demonstrate the search algorithm's effectiveness under lower computational resources with better clarity.

\begin{table}[!htb]
    \centering
    \begin{tabular}[b]{c}
        \begin{tabular}{c|cccc}
        \toprule
         Method & Cost (MB) & Panc. & Tumor & Avg. \\
        \midrule
        \multirow{3}{*}{\small{DiNTS~\cite{DBLP:conf/cvpr/He0RZX21}}}
        & 5,067
        & 77.94
        & 48.07
        & 63.00\\
        & 7,239
        & 80.20
        & 52.25
        & 66.23\\
        & 8,802
        & 80.06
        & 52.53
        & 66.29\\
        \midrule
        \multirow{3}{*}{\small{HyperSegNAS}}
        & 5,358
        & 79.92 
        & 52.44
        & 66.18\\
        & 7,168
        & 80.10
        & 53.47
        & 66.79\\
        & 9,173
        & 79.98
        & 54.88
        & 67.43\\
        \bottomrule
        \end{tabular}\\
    \end{tabular}
    \caption{Performance comparison for architectures found under different compute constraints, based on five-fold cross-validation\protect\footnotemark on the Pancreas dataset.}
    \label{tab:model_compare}
    \vspace{-1em}
\end{table}

\begin{table*}[!htb]
    \setlength{\tabcolsep}{2pt}
    \centering
    \begin{tabular}[b]{c}
        \begin{tabular}{l|ccc|ccc|cc|ccc|ccc}
        \toprule
            &\multicolumn{6}{c|}{Pancreas and Tumor}&\multicolumn{2}{c|}{Lung Tumor}&\multicolumn{6}{c}{Hepatic Vessel and Tumor}\\
        \midrule
         \multicolumn{1}{c|}{Method} & DSC1 & DSC2& Avg. & NSD1 & NSD2 & Avg.& DSC1 & NSD1 & DSC1 & DSC2& Avg. & NSD1 & NSD2 & Avg. \\
        \midrule
        Kim et al~\cite{DBLP:conf/miccai/KimKLBKCYK19} & 80.61 &51.75& 66.18& 95.83& 73.09&84.46& 63.10 &62.51&62.34& 68.63& 65.49& 83.22& 78.43& 80.83\\
        nnUNet~\cite{Isensee2021}
        & \textbf{81.64}& 52.78& 67.21& 96.14& 71.47& 83.81&73.97& 76.02&\textbf{66.46}& \underline{71.78}& \textbf{69.12} &\textbf{84.43} &80.72 &\underline{82.58}
        \\
        C2FNAS~\cite{DBLP:conf/cvpr/YuYRBZYX20}
        & 80.76& 54.41& 67.59 &96.16& 75.58& 85.87&70.44& 72.22& 64.30& 71.00& 67.65& 83.78 &80.66& 82.22\\
        DiNTS~\cite{DBLP:conf/cvpr/He0RZX21}
         & \underline{81.02} & \underline{55.35} & \underline{68.19} & \underline{96.26} &\underline{75.90} &\underline{86.08}& \underline{74.75} & \underline{77.02}&\underline{64.50} &71.76& 68.13& 83.98& \underline{81.03}& 82.51\\
        HyperSegNAS
        & 80.99& \textbf{56.16}& \textbf{68.58}& \textbf{96.30} & \textbf{77.68} & \textbf{86.99}&\textbf{76.72}& \textbf{79.67}& 64.35&	\textbf{72.10}&	\underline{68.23}&\underline{84.08}	&\textbf{81.14}&\textbf{82.61}\\
        \bottomrule
        \end{tabular}\\
    \end{tabular}        
    \begin{tabular}[b]{c}
        \begin{tabular}{l|cccc|cccc|ccc|ccc}
        \toprule
        &\multicolumn{8}{c|}{Brain Tumor} &\multicolumn{6}{c}{Hippocampus}\\
        \midrule
         \multicolumn{1}{c|}{Method} & DSC1 & DSC2 & DSC3 & Avg.& NSD1 & NSD2 & NSD3 & Avg. &DSC1 & DSC2& Avg. & NSD1 & NSD2 & Avg.\\
        \midrule
        Kim et al~\cite{DBLP:conf/miccai/KimKLBKCYK19} &67.40& 45.75& 68.26& 60.47& 86.65& 72.03& 90.28& 82.99& 90.11& 88.72& 89.42& 97.77& \underline{97.73}& \underline{97.75}\\
        nnUNet~\cite{Isensee2021}
        & 68.04 & 46.81& 68.46& 61.10& 87.51& 72.47& 90.78& 83.59 &
        \textbf{90.23}&\textbf{88.69} & \textbf{89.46}& \underline{97.79}&	97.53& 97.66\\
        C2FNAS~\cite{DBLP:conf/cvpr/YuYRBZYX20}
        & 67.62& 48.60& 69.72& 61.98& 87.61& 72.87& 91.16& 83.88&
        89.37 &87.96& 88.67& 97.27 &97.35 &97.31\\
        DiNTS~\cite{DBLP:conf/cvpr/He0RZX21}
        & \underline{69.28}& \underline{48.65} & \underline{69.75}& \underline{62.56}& \underline{89.33}& \textbf{73.16}& \underline{91.69}& \textbf{84.73}&
        89.91 &88.41 &89.16 &97.76 &97.56 &97.66\\
        HyperSegNAS
        & \textbf{69.32}& \textbf{49.11}& \textbf{70.01}& \textbf{62.81}& \textbf{89.76}& \underline{72.58} & \textbf{91.74} & \underline{84.69}&
        \underline{90.21}&	\underline{88.47}&	\underline{89.34} &\textbf{98.11}&\textbf{97.7}&\textbf{97.91}\\
        \bottomrule
        \end{tabular}\\
    \end{tabular}
    \caption{Dice-Sørensen score (DSC) and Normalised Surface Distance (NSD) performance on the Brain Tumor (MRI), Hippocampus (MRI), Lung Tumor (CT), Pancreas and Tumor (CT), Hepatic Vessel and Tumor (CT) datasets from MSD. All scores are from the current MSD leaderboard. The best and second best performances are \textbf{bold} and \underline{underlined}.}
    \label{tab:msd_results}
    \vspace{-1em}
\end{table*}

DiNTS~\cite{DBLP:conf/cvpr/He0RZX21} yields a low-cost architecture that downsamples all features to the lowest resolutions. Such an architecture differs from conventional designs, where multi-scale features are desired for segmenting objects of various sizes; as a result, the performance of this architecture is worse than HyperSegNAS's low-cost architecture. We attribute DiNTS's counter-intuitive results to differentiable NAS formulation. In DiNTS, an edge can have a near-zero importance ($\eta^i_j$ in~\cite{DBLP:conf/cvpr/He0RZX21}) and estimated memory cost while still propagating its features forward, leading to a significant discrepancy between search and deployment. In comparison, HyperSegNAS evaluates on discrete representations, and reflects its performance in deployment more closely. In fact, \emph{our smallest architecture only performs marginally worse than the largest architecture from DiNTS} in Table~\ref{tab:model_compare}, demonstrating HyperSegNAS's ability to truly adapt to different computing constraints.
\footnotetext{We thank the authors from ~\cite{DBLP:conf/cvpr/He0RZX21} for providing their 5-fold data splits for fair comparisons.}

Secondly, skip connections are used extensively in our architectures. This shows the advantage in occupying as much of the search space as possible, as these connections make the original features more available to other parts of the network with little cost. Such an observation generally agrees with previous works like ResNet~\cite{DBLP:conf/cvpr/HeZRS16} and DenseNet~\cite{DBLP:conf/cvpr/HuangLMW17}, where features are shuffled through skip connections to achieve less information loss and improved representations. In particular, we find parallel skip connections to be very efficient, as they introduce very little computing overhead and consistently improve architecture performance according to evaluations based on the super-net. This is in agreement with the design of U-Net~\cite{DBLP:conf/miccai/RonnebergerFB15}, where parallel skip connections are also used extensively.
\begin{figure}[!htb]
    \setlength{\abovecaptionskip}{3pt}
    \setlength{\tabcolsep}{0.2pt}
    \vspace{-1em}
    \begin{tabular}[b]{cc}
        \begin{subfigure}[b]{0.5\linewidth}
            \includegraphics[width=\textwidth]{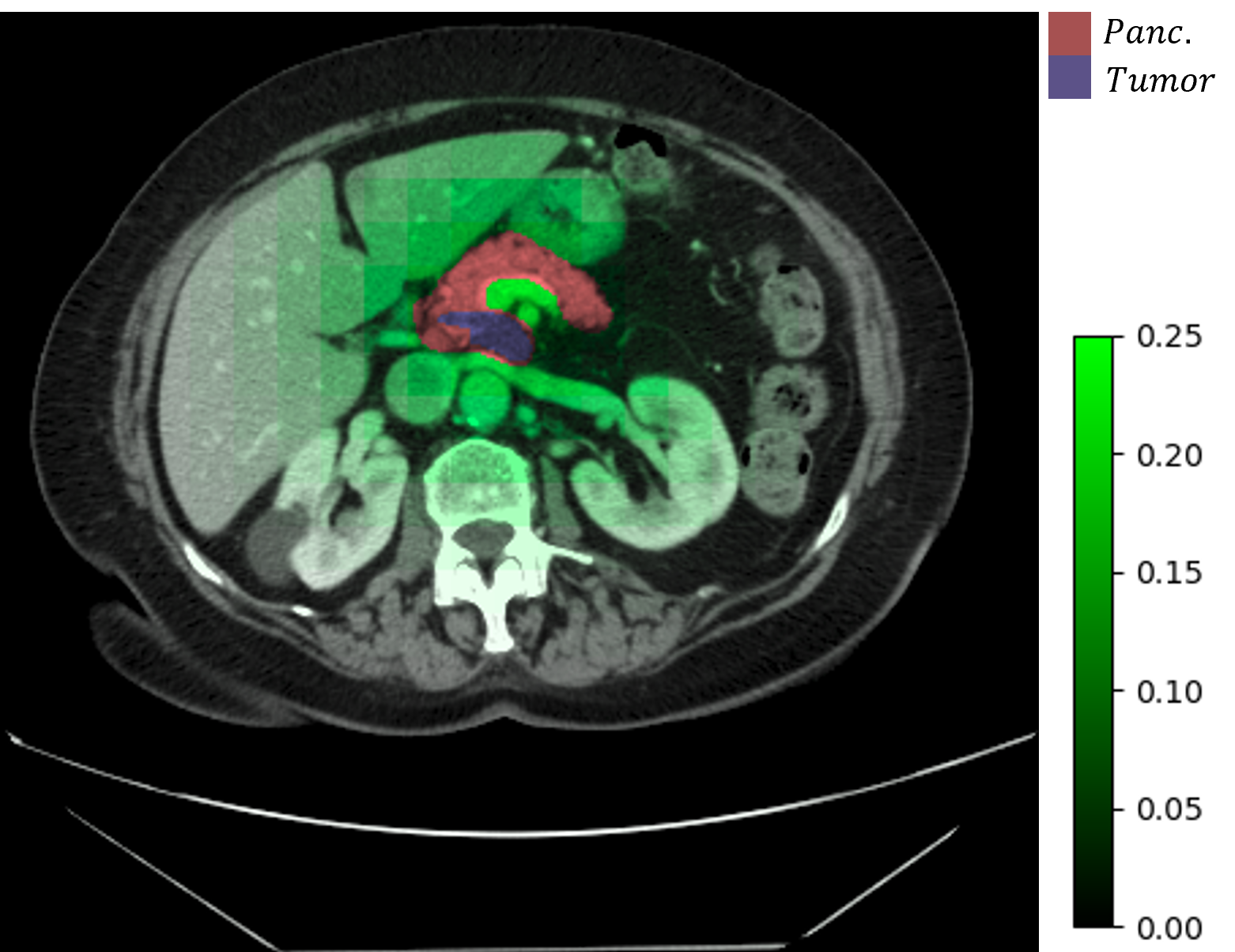}
            \caption{$\Delta\omega^{a}_{\mathcal{H}_\theta}(I)$ and labels}
        \end{subfigure} &
        \begin{subfigure}[b]{0.5\linewidth}
            \includegraphics[width=\textwidth]{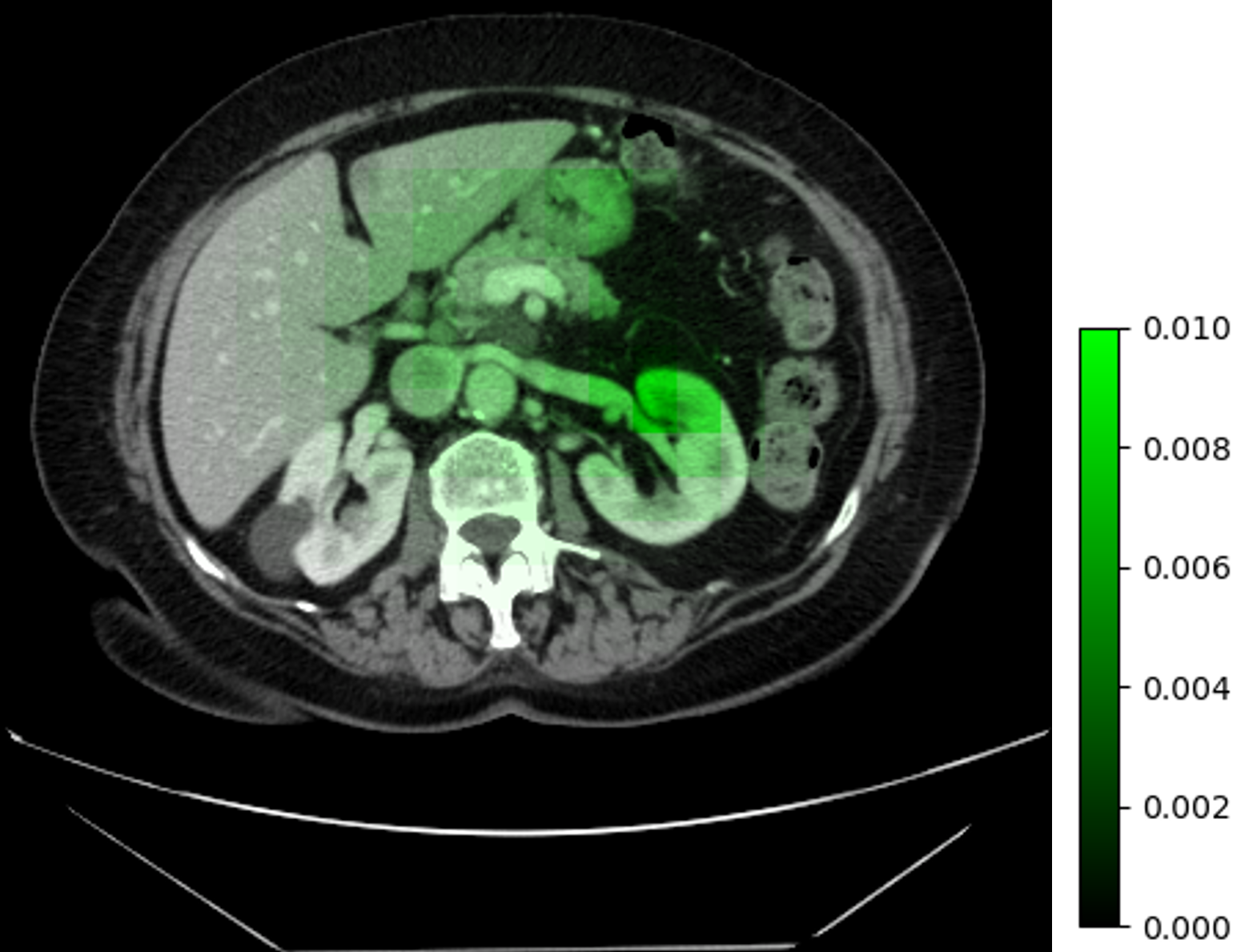}
            \caption{$\Delta\omega^{I}_{\mathcal{H}_\theta}(a_i, a_j)$}
            \label{architect_delta}
        \end{subfigure} 
    \end{tabular}
    \caption{Visualization of changing $\omega^{a,I}_{\mathcal{H}_\theta}$ depends on different image or architecture inputs. We use our 5GB and 7GB architectures as $a_i$ and $a_j$ in \ref{architect_delta}. Results are aggregated based on sliding-window averaging with a stride size of sixteen.}
    \label{fig:visualization}
    \vspace{-1.25em}
\end{figure}

\textbf{HyperNet Analysis.} We provide some insights into the behavior of the HyperNet $\mathcal{H}$ and its output $\omega^{a,I}_{\mathcal{H}_\theta}$. While it is difficult to interpret the channel-wise weights individually, we can observe how $\omega^{a,I}_{\mathcal{H}_\theta}$ changes with different inputs. To do so, we fix the architecture vector $l_{\textrm{arch}}$ and provide $\mathcal{H}$ with $l_{\textrm{image}}$ generated from different patches. Similarly, we can fix $l_{\textrm{image}}$ and change $l_{\textrm{arch}}$ to observe the differences based on architectures. We observe that the majority of the patches yields very similar $\omega^{a,I}_{\mathcal{H}_\theta}$, while a few patches have very different weights. To quantify, we define two metric:
 \begin{align}
 \Delta\omega^{a}_{\mathcal{H}_\theta}(I)=\lVert\omega^{a,I}_{\mathcal{H}_\theta}-\overline{\omega}^{a}_{\mathcal{H}_\theta}\rVert_2\\
 \Delta\omega^{I}_{\mathcal{H}_\theta}(a_i, a_j)=\lVert\omega^{a_i,I}_{\mathcal{H}_\theta}-\omega^{a_j,I}_{\mathcal{H}_\theta}\rVert_2, 
 \end{align}

where $\Delta\omega^{a}_{\mathcal{H}_\theta}(I)$ measures the $\mathcal{L}_{2}$ distance between $\omega^{a,I}_{\mathcal{H}}$, which is from a specific patch $I$, and $\overline{\omega}^{a}_{\mathcal{H}}$, which is the average weights over all patches in a given volume; $\Delta\omega^{I}_{\mathcal{H}_\theta}(a_i, a_j)$ measures the difference between weights generated from architecture $a_i$ and $a_j$ on the same patch.

We visualize these two metrics on a sample CT volume in Fig.~\ref{fig:visualization}, along with the segmentation labels. For patches that are not relevant to the labels, $\mathcal{H}$ generates very similar $\omega^{a,I}_{\mathcal{H}_\theta}$; on the other hand, patches that contain labels yield significantly different weights. Furthermore, $\Delta\omega^{I}_{\mathcal{H}_\theta}(a_i, a_j)$ is also significant on foreground patches, and minimum on background patches. This suggests that $\mathcal{M}$ and $\mathcal{H}$ implicitly partitions potential foreground regions, which is a strategy similar to foreground oversampling used in methods like nnU-Net~\cite{Isensee2021}. While our method does not employ explicit foreground oversampling, the HyperNet design appears to automate such a strategy. For more details on the visualization of $\omega^{a,I}_{\mathcal{H}_\theta}$, please refer to the Supplemental Material.

\subsection{Quantitative Evaluation on MSD}

We test our largest architecture on five tasks in the MSD challenge - the Pancreas and Tumor, Lung Tumor, Hepatic Vessel and Tumor, Brain Tumour, Hippocampus datasets. We select these tasks as a simplified and representative set of MSD to make experiments more efficient, as these tasks cover 80$\%$ of all data in MSD and include multiple organs and imaging modalities. The DSC and Normalised Surface Distance (NSD) scores are reported from the MSD leaderboard, and listed in Table~\ref{tab:msd_results}. HyperSegNAS' architecture shows an overall improvement on the Pancreas dataset compared to previous NAS-based methods, particularly on Pancreatic Tumour (DSC2), which is the more challenging and clinically relevant label. For tasks of the same CT modality, our architecture significantly outperforms all baselines in Lung Tumour segmentation, and is $\sim$2 better than the previous SOTA from DiNTS in DSC. General improvements can also be found in MRI segmentation tasks against other methods, showing a generality to our architecture. 

We note that while nnU-Net~\cite{li2018hdenseunet} outperforms HyperSegNAS on some metrics in Hepatic Vessel and Hippocampus, it achieves so by (1) manually searching for the best hyper-parameters and (2) performing ensemble over different variations of U-Net. Specifically, to generate results for a single task, nnU-Net trains \emph{four} architecture configurations each with five-fold cross validation, which leads to an ensemble of \emph{twenty} models. HyperSegNAS only performs ensemble over five models from cross validation, uses no task-specific hyper-parameter tuning, and obtains similar if not better performances compared to nnU-Net overall. 



\subsection{Limitation and Future Directions}
This work mainly focuses on obtaining general segmentation architectures. Problem specific factors like hyper-parameter selection and data augmentation methods are also important, particularly on the smaller datasets in MSD which we do not explore. The use of 2D models or 2D operators can also be beneficial to tasks where the volume resolution within-slice and between-slice is very different.

As the topology search space for segmentation is expansive, training a super-net requires high compute cost. While ability to compare architectures yields useful insights, e.g., on the incorporation of skip connections, we plan to investigate ways to partition the segmentation search space into more manageable pieces and improve search efficiency.

\section{Conclusions}
In this work, we present HyperSegNAS, a novel NAS algorithm for medical image segmentation. HyperSegNAS finds well-performing segmentation architectures by estimating their performances through a trained super-net. To address the complexity in the training process due to a large and unordered segmentation search space, we propose to use a HyperNet called Meta-Assistant Network (MAN) to improve learning. MAN incorporates high level information on the deploying architecture and the input image to output channel-wise weights, which modify the inference process for better performance. To allow fair evaluations during the searching stage, HyperSegNAS then removes MAN through a annealing process. We perform extensive experiments and show that our method can effectively adapt to different compute constraints both in terms of precision in matching the constraints and segmentation performance. Specifically, our low-compute architecture significantly outperforms DiNTS's architecture on the Pancreas dataset. We also test our large-compute architecture on five representative tasks in MSD, and achieve SOTA performances respectively. Our analysis of architecture performance estimations show that the inclusion of skip connections is crucial to finding well-performing architectures, which agrees with the intuition from a history of manual architecture designs and is not reflected in previous methods.

{\small
\bibliographystyle{ieee_fullname}
\bibliography{main}
}

\end{document}